\documentclass[referee]{raa}

\usepackage{graphicx,times}
\usepackage{natbib}
\usepackage{amssymb,amsmath}
\bibpunct{(}{)}{;}{a}{}{,}

\usepackage[pagebackref=true]{hyperref}

\begin{document}

   \title{Photometric properties and stellar parameters of the rapidly rotating magnetic early-B star HD\;345439}

 \volnopage{ {\bf 20XX} Vol.\ {\bf X} No. {\bf XX}, 000--000}
   \setcounter{page}{1}

   \author{Dong-Xiang Shen\inst{1}, Jin-Zhong Liu\inst{2}, Chun-Hua Zhu\inst{1}, Guo-Liang L{\"u}\inst{1}, Yu Zhang\inst{2}, Cheng-Long L{\"u}\inst{2}, Hao-Zhi Wang\inst{2}, Lei Li\inst{1}, Xi-Zhen Lu\inst{1}, Jin-Long Yu\inst{1}, Abdurepqet Rustem\inst{1}
   }
% Here is an example of three authors come from different institutes.
% For single author or all the authors from an institute, use "\inst{}" only

   \institute{School of Physical Science and Technology, Xinjiang University, Urumqi 830046, China; {\it chunhuazhu@sina.cn}\\
% Please give the E-mail address of the author, to whom future correspondence and offprint requests will be sent.
        \and
            Xinjiang Astronomical Observatory, National Astronomical Observatories, Chinese Academy of Sciences, Urumqi 830000, China; {\it liujinzh@xao.ac.cn}\\
\vs \no
   {\small Received 20XX Month Day; accepted 20XX Month Day}}

\abstract{We first present the multicolor photometry results of the rapidly rotating magnetic star
HD\,345439 using the Nanshan One-meter Wide-field Telescope. From the photometric observations,
we derive a rotational period of $0.7699\pm0.0014\;$day. The light curves of HD\,345439 are dominated by the double asymmetric S-wave feature that arises from the magnetic clouds. Pulsating behaviors are not observed in Sector 41 of the Transiting Exoplanet Survey Satellite. No evidence is found of the occurrence of centrifugal breakout events neither in the residual ﬂux nor in the systematic variations at the extremum of the light curve. Based on the hypothesis of the Rigidly Rotating Magnetosphere model, we restrict the magnetic obliquity angle {$\beta$} and the rotational inclination angle $i$ so that they satisfy the approximate relation {$\beta + i \approx 105^{\circ}$}. The colour excess, extinction, and luminosity are determined to be $E_{(B-V)}=0.745\pm0.016\,$mag, $A_{V}=2.31\pm0.05\,$mag, and $\rm log\,(L/L_{\odot})=3.82\pm0.1\;$dex, respectively. Furthermore, we derive the effective temperature as $\; T$$\rm _{eff}=22\pm1\;$kK and the surface gravity as log$\,g=4.00\pm0.22$. The mass$\,M=7.24_{-1.24}^{+1.75}\rm\,M_{\odot}$, radius$\,R=4.44_{-1.93}^{+2.68}\rm\,R_{\odot}$, and age$\rm\,\tau_{age}=23.62\,_{-21.97}^{+4.24}\,$Myr are estimated from the Hertzsprung--Russell Diagram.
\keywords{stars: fundamental parameters ---  stars: early-type --- stars: magnetic fields --- stars: rotation ---  stars: chemically peculiar ---  stars: individual (HD\,345439)
}
}

   \authorrunning{D.-X. Shen et al. }            %author_head in even pages
   \titlerunning{Photometry and stellar parameters of HD\,345439}  % title_head in odd pages
   \maketitle

%________________________________________________ sections below
%
\section{INTRODUCTION}
Since pioneering detections of the strong magnetic ﬁelds in the chemically peculiar Ap/Bp stars (\citealt{1947ApJ,1980ApJS...42..421B}), new generations of spectropolarimeters
have detected large-scale, systematic magnetic fields in numerous massive stars \citep[e.g.,][]{2003A&A403645B,2004MNRAS351826P,2011A&A535A25B,2012MNRAS4192459W,
2012MNRAS4251278W,2015A&A574A20,2019A&A626A94G}. In recent years,
the Magnetism in Massive Stars \citep[MiMeS;][]{2009IAUS..259..333W} and
B fields in OB stars \citep[BOB;][]{2014A&A564L10} projects have detected dozens of new magnetic OB
stars successively; a proportion of these magnetic stars have also been revealed using high-resolution spectroscopy \citep[e.g.,][]{ 2006A&A...451..293W,2012MNRAS.425.1278W,2015MNRAS.447.2551W}. The observations indicate that magnetic stars exhibit unique
phenomena, such as cyclic variability in line profile morphologies \citep[e.g.,][]{2010MNRAS.407.2383F,2017MNRAS.472..400H,2017MNRAS.468.3985S},
 X-ray emission \citep[e.g.,][]{2008A&A...483..857S,2011AN....332..988O}, and light curve variations \citep[e.g.,][]{2014A&A...562A.143F,2019A&A...626A..94G,2021MNRAS.507.1283S}.
The behaviors that magnetic fields cause in massive stars, for instance, magnetic field evolution, mass-loss quenching, and magnetic braking, are of utmost importance to understanding the evolution of magnetic stars.

Being a typical helium-rich massive star, HD\;345439 shows extremely strong, large-scale magnetic
fields and fast rotation \citep{2014ApJ...784L..30E}. The longitudinal magnetic fields of HD 345439 have a periodic variability with an amplitude of 2-3 kG, and the polar magnetic strength reaches 10 kG \citep{2017MNRAS.467L..81H}. To date, only several analogs have been discovered: HR\;7355 \citep{2008A&A...482..255R},
HR\;5907 \citep{2012MNRAS.419.1610G}, HD\;23478 \citep{2015ApJ...811L..26W}, {$\sigma$}\;Ori\;E \citep{1978ApJ...224L...5L}, HD\;176582 \citep{2011AJ....141..169B},
{$\delta$}\;Ori\;C \citep{2010MNRAS.401.2739L}, HD\;35502 \citep{2016MNRAS.460.1811S},
and HD\;164492C \citep{2017MNRAS.467..437G,2017MNRAS.465.2517W}. The observations indicate that strong magnetic
fields may contribute to material accumulation and the formation of a high-density disk \citep{2013ApJ...766L...9C,2015MNRAS.451.2015O}. On the other hand, the magnetic field seems to result in an enhanced He absorption and an inhomogeneous spatial distribution of the elements in the photosphere \citep[e.g.,][]{2012MNRAS.419.1610G,2013MNRAS.429..177R}. Other physical manifestations of the magnetosphere were detected in observations of X-ray, ultraviolet\,(UV), and radio observations \citep[e.g.,][]{2018MNRAS.476..562L}.

Recently, the Rigidly Rotating Magnetosphere\;(RRM)\;model has been developed to explain the variability of magnetic stars \citep{2005MNRAS.357..251T,2005ApJ...630L..81T}. The RRM model provides an empirical picture according to which plasma is predominantly subjected to gravity and the centrifugal force. Furthermore, the plasma originated from stars is constrained to move along the magnetic ﬁeld lines. The gravitational force prompts the material to infall toward the star, while the centrifugal force acts in the opposite manner. Based on the assumptions of the RRM model, there is a point for each magnetic ﬁeld line where the gravitational plus centrifugal potential reaches the minimum value required for causing the material accumulation. The disk consisting of the accumulated material produces the eclipse morphology in the light curves of {$\sigma$} Ori E type stars \citep{2015MNRAS.451.2015O}. Until now, only {$\sigma$} Ori E has been characterized using photometry, spectropolarimetry, and spectrometry. Here, we aim to provide more detailed characteristics of the rapidly rotating magnetic early-type star HD\,345439 using NOWT photometric observations.

We introduce the observations of the photometric and spectrometric data in Section 2 of this article. In Section 3, we determine the rotational period, pulsating behavior, variations in the light curves, centrifugal breakout, and stellar parameters of HD\,345439 from the collected datasets and the theoretical model. In Section 4, we discuss the results derived from the observations. Finally, the conclusions are provided in Section 5.

\section{OBSERVATIONS AND DATA REDUCTIONS}
\subsection{Multicolor photometric observations}
The multicolor photometric measurements of HD\,345439 were performed using the Nanshan One-meter Wide-field Telescope \citep[NOWT;][]{2020RAA....20..211B,2021RAA....21..124S} of the Xinjiang Astronomical Observatory\;(XAO), Chinese Academy of Sciences\;(CAS). NOWT is an alt-azimuth mount reflector telescope at the Nanshan in Xinjiang. NOWT is equipped with a 4096$\times$4136-pixel CCD and a Johnson-Cousins $UBVRI$ filter system. The CCD provides a field of view\;(FoV)\;of $1.3\times1.3$ degree$^{2}$, and the per-pixel resolution is 1.28$''$. The photometric observations were performed over a period of 12 nights from October 13 to November 12, 2020, and 2457 frames of HD\,345439 were obtained. During the observations, when a single exposure of one band was finished, the ﬁlter wheel rotated to the next band. The FoV of the CCD readout was set as $42.67\times42.67$ arcmin$^{2}$ to ensure the efficiency of the observations, and the scan rate mode was set to medium. The observations were supplemented with a series of 10 bias and 3 flat-field exposures per filter every night to calibrate the CCD instrumental signature. The observing log is summarized in Table\;\ref{tab:observation-log}.

\begin{figure}
\centering
\includegraphics[scale=0.35]{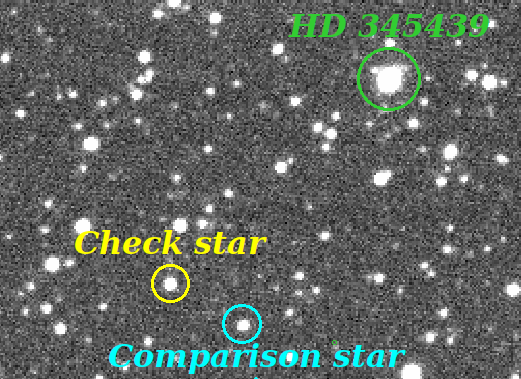}
\caption{Part of one observation image, where HD 345439, the comparison star, and the check star are indicated.}
\label{fig:observations_image}
\end{figure}

\begin{figure}
\centering
\includegraphics[scale=0.5]{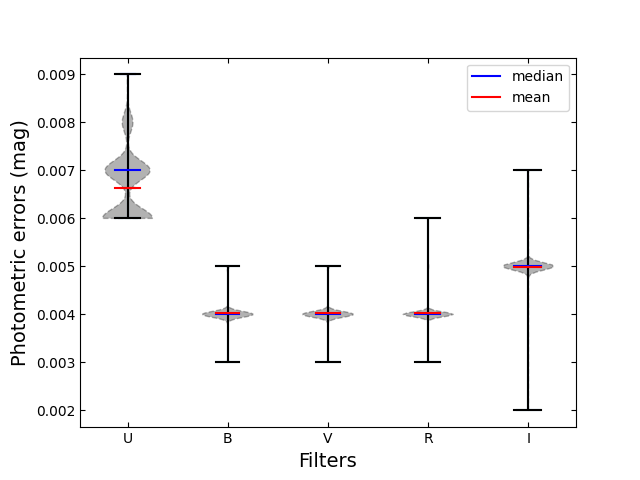}
\caption{Photometric error distribution of each multicolor photometry filter. The error bars represent the maximum and minimum values of the photometric errors. The gray areas represent the kernel density distribution of the photometric errors. The red and blue segments represent the mean and median values of the photometric errors, respectively.}
\label{fig:photometric_error}
\end{figure}

The data reductions, including bias subtraction, ﬂat-ﬁelding, and illumination correction, were carried out using the CCDPROC task of the Image Reduction and Analysis Facility\;(IRAF\footnote{IRAF is distributed by the National Optical Astronomy Observatories, which are operated by the Association of Universities for Research in Astronomy, Inc., under cooperative agreement with the National Science Foundation.}) software. Part of one observation image is shown in Fig.\;\ref{fig:observations_image}, where HD\;345439, the comparison star, and the check star are indicated. Differential photometry was carried out by using the standard aperture photometric package of the IRAF APPHOT task. The photometric error distribution is shown in Fig.\;\ref{fig:photometric_error}. Although the photometric error in the $U$ band is considerably higher than that in the other bands, the overall multicolor photometry process is stable.
\begin{table}
\caption[]{Summary of Photometric Observations with NOWT\label{tab:observation-log}}
\small
\begin{tabular}{ccccccc}
\hline\noalign{\smallskip}
\hline\noalign{\smallskip}
Date& U &  B&  V & R &  I& Weather \\
&  Exp. 45s & Exp. 10s & Exp. 7s &Exp. 7s & Exp. 8s& \\
\hline\noalign{\smallskip}
2020 Otc 13 & 60 & 59 & 58 & 57 & 57 &P.\\
2020 Otc 14 & 59 & 58 & 59 & 59 & 57 &P.\\
2020 Otc 15 & 40 & 36 & 36 & 35 & 35 &P.\\
2020 Otc 16 & 63 & 67 & 63 & 62 & 62 &P.\\
2020 Otc 17 & 55 & 56 & 63 & 52 & 52 &P.\\
2020 Otc 18 & 35 & 36 & 36 & 36 & 36 &P.\\
2020 Otc 19 & 55 & 55 & 55 & 55 & 55 &P.\\
2020 Otc 20 & 47 & 47 & 46 & 46 & 45 &P.\\
2020 Otc 21 & 43 & 41 & 42 & 41 & 41 &P.\\
2020 Nov 8  &  6 & 6  & 6  & 5  & 6  &C.\\
2020 Nov 11 & 19 & 19 & 19 & 19 & 19 &C.\\
2020 Nov 12 & 17 & 15 & 16 & 16 & 16 &C.\\
  \noalign{\smallskip}\hline
\end{tabular}
\tablecomments{0.86\textwidth}{This table summarizes the situations of photometric observations of NOWT. The number of frames for different filters in each night are listed, respectively. The exposure times are listed under the filter name. In the last column, we marked the weather of each observing night, P. represents the relatively good night, and the Full Width at Half Maximum (FWHM) of the HD\;345439 less than 2.8, and C. represents the cloudy night, with the FWHM of HD\;345439 less than 3.5 and larger than 2.8.}
\end{table}

\subsection{Archival data}
The archival photometric data of HD\;345439 employed in this work were obtained from the All-Sky Automated Survey for Supernovae \citep[ASAS-SN;][]{2017PASP..129j4502K,2018MNRAS.477.3145J,2019MNRAS.487.5932P}, the Super Wide Angle Search for Planets \citep[SuperWASP;][]{2010A&A...520L..10B}, and the Transiting Exoplanet Survey Satellite \cite[TESS;][]{2015JATIS...1a4003R}.
The ASAS-SN was the ﬁrst project to routinely survey the entire visible sky in the V band, reaching a depth of roughly 17 mag. As of the mid-2017, ASAS-SN consists of two observing stations, namely the Haleakala Observatory (Hawaii) and the Cerro Tololo International Observatory (CTIO, Chile) sites. By the end of 2017, ASAS-SN expanded to comprise ﬁve observing stations with the addition of a second unit at the CTIO and two more units at the McDonald Observatory (Texas) and the South African Astrophysical Observatory (SAAO, Sutherland, South Africa). Two observational datasets were obtained from the ASAS-SN Photometry Database\footnote{https://asas-sn.osu.edu/variables} using different sources. In the ﬁrst observational dataset, ASAS-SN observed HD\;345439 using the APOGEE input catalog; 349 photometric data points were obtained from February 2015 to April 2018 using bc and bd cameras. In the second observational dataset, the observations were conducted using the ASAS-SN DR9 source and the bc camera and consisted of 192 epochs from February 2015 to October 2018. SuperWASP is an extra-solar planet detection program employing identical robotic telescopes. The telescopes are equipped with a 2048$\times$2048 thinned CCD with a pixel size of 13.5 $\mu$m. SuperWASP carried out observations of HD\,345439 over 67 nights in 2007 and provided 979 points of photometric observations in a broadband ﬁlter (bandpass from 400 to 700 nm). TESS is a NASA Astrophysics Explorer mission. The satellite is equipped with four identical wide-ﬁeld cameras, which together can monitor the sky. TESS conducted photometric observations of HD\,345439 in Sector 41 with a 2-min cadence mode, providing the most complete light curve of the star. The photometric data of TESS was reduced using the lightkurve package\;(hereafter denoted as LK)\footnote{http://docs.lightkurve.org/}.

The Large Sky Area Multi-Object Fiber Spectroscopic Telescope \citep[LAMOST;][]{2012RAA....12.1197C} is a reﬂecting Schmidt telescope located at the Xinglong station of the National Astronomical Observatories, Chinese Academy of Sciences. The mean aperture of the LAMOST is 4.3 m, and the corresponding FoV is 5 degree in diameter \citep{2015RAA....15.1095L}. A Low-resolution spectrum of HD\,345439 was obtained in the LAMSOT Low-Resolution Survey Data Release 7\;(LAMOST-LRS DR7)\footnote{http://dr7.lamost.org/}.

\begin{figure}
\begin{minipage}[t]{0.25\linewidth}
\centering
\includegraphics[width=6in]{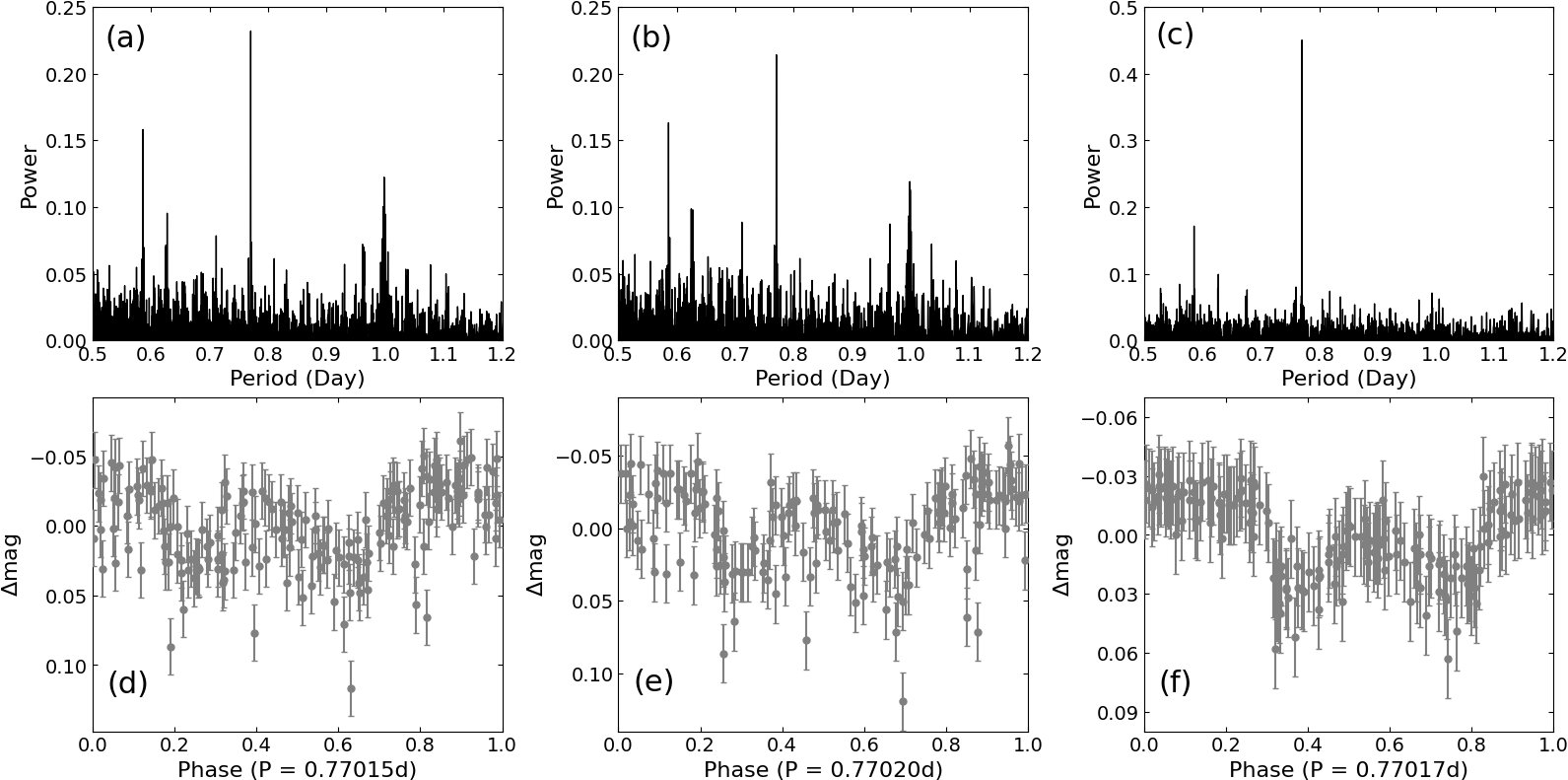}
\end{minipage}\\
\begin{minipage}[t]{0.25\linewidth}
\centering
\includegraphics[width=5.5in]{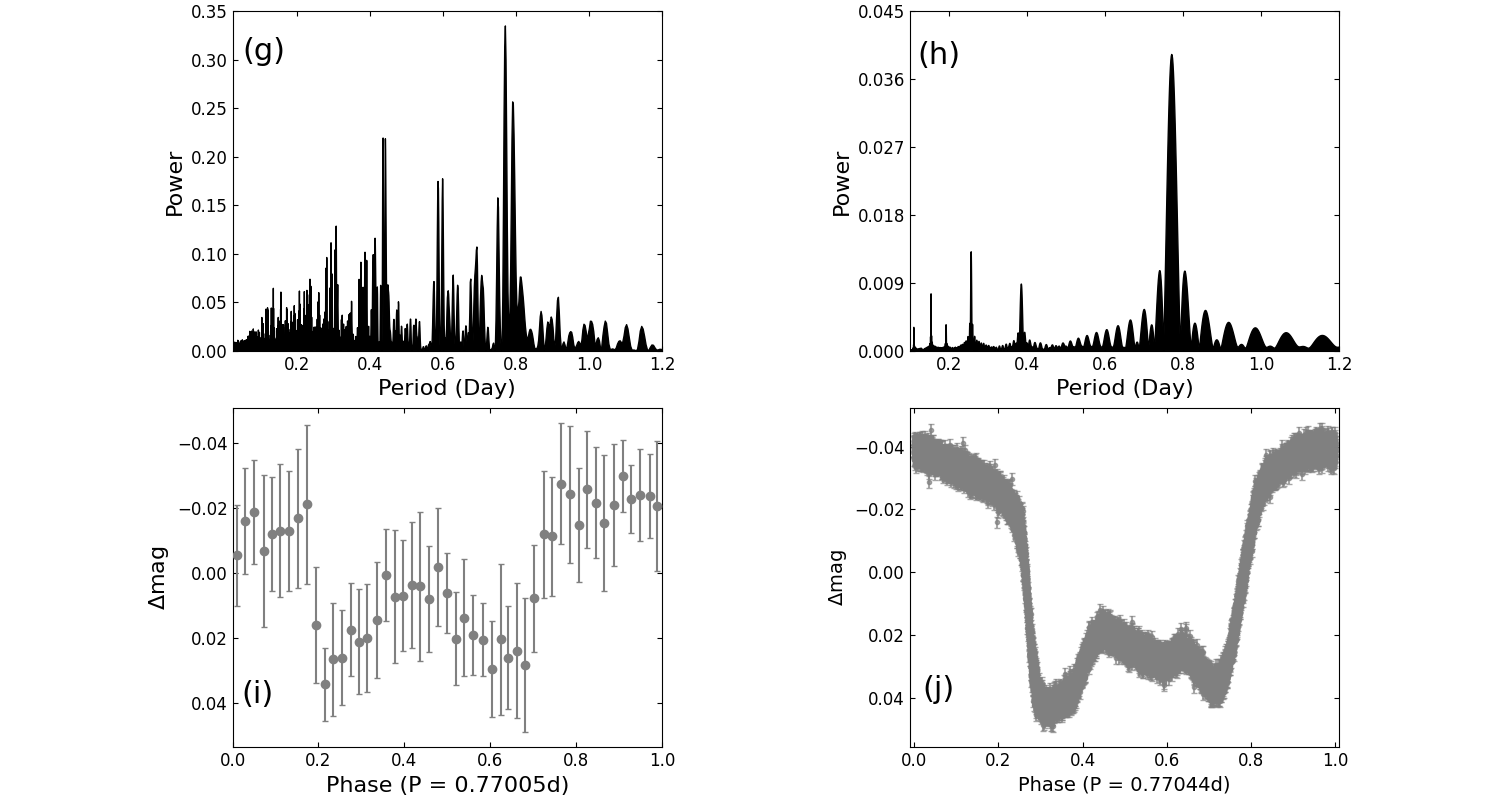}
\end{minipage}%
\caption{Periodograms (upper panels) and corresponding phase diagrams (bottom panels) of each archival dataset of HD\;345439. Panel (a) shows the periodic spectrum of the observations of ASAS-SN with the source of ASAS-SN DR9. Panel (b) shows the periodic spectrum of the ASAS-SN observations (with the APOGEE source) using the bc camera, while panel (c) illustrates the periodic spectrum of the ASAS-SN observations (with the APOGEE source) using the bd camera. The homologous phase diagrams are listed at the bottom of each periodic spectrum. The same diagrams for SuperWASP and TESS data are shown in panels (g) to (j).}
\label{fig:periodogram}
\end{figure}

\section{RESULTS}
\subsection{Rotational period}
To ﬁt the He\;{\uppercase\expandafter{\romannumeral1}} line proﬁles, \cite{2014ApJ...784L..30E} hypothesized that the rotational velocity $v$sin$i$ has a value of $270\pm20$ km s$^{-1}$.
\cite{2015ApJ...811L..26W} derived a rotational period of 0.7701 day from the KELT, SuperWASP, and ASAS surveys. In this work, we determine the rotational period of HD\;345439 by analyzing the new archival data and new photometric observations. The Lomb-Scargle method\;\citep{1976Ap&SS..39..447L,1982ApJ...263..835S} provided by Astropy \citep{2013A&A...558A..33A,2018AJ....156..123A} was used to detect the periodic signals in the unevenly spaced SuperWASP and ASAS-SN observations. The PERIOD04 software \citep{2005CoAst.146...53L} was used to detect the periodic signature in both the TESS and NOWT photometric observations. The validity of the frequencies that detected in the TESS and NOWT photometric observations has followed the example of \cite{1993A&A...271..482B}. The period-power spectra and corresponding light curves of the archival data are shown in Fig.~\ref{fig:periodogram}.

The Lomb–Scargle periodogram of ASAS-SN using the ASAS-SN DR9 source shows that the rotational period exhibits a distinct signature at $0.7701\pm 0.0002$ day. The value of $0.7702\pm0.0001$ day was derived from the ASAS-SN observations using the APOGEE source with the bc camera. A consistent periodic signal with a value of $0.77017\pm0.00013$ day appears also in the ASAS-SN measurements using the bd camera. The SuperWASP photometric data also reveals a rotational period of $0.77\pm0.0042$ day. The TESS observations indicate a rotational period of around 0.77044 $\pm$ 0.00009 day, and this result is statistically consistent with previous results. We summarized the rotational periods obtained from the different datasets in Table~\ref{tab:period_summarized}.

\begin{table}
\caption{Rotational period in different Survey}
\label{tab:period_summarized}
\begin{tabular}{cccccc}
\hline\noalign{\smallskip}
\hline\noalign{\smallskip}
Survey& Filter& $f_{rot}$  & Period &  Period error &Remark\\
      &       &(day$^{-1}$)&  (day) &    (day)&\\
\hline\noalign{\smallskip}
ASAS-SN& V    & 1.2985327 & 0.7701   &  0.0002    &bc,APOGEE\\
      &  V    & 1.2983641 & 0.7702   &  0.0001    &bd,APOGEE\\
      &  V    & 1.2984146 & 0.77017  &  0.00013   &bd,ASAS-SN DR9\\
NOWT  &  U    & 1.3017526 & 0.7682053&  0.0027970 &\\
      &  B    & 1.2991486 & 0.7697400&  0.0019904 &\\
      &  V    & 1.3008506 & 0.7687349&  0.0023462 &\\
      &  R    & 1.2988267 & 0.7699334&  0.0024346 &\\
      &  I    & 1.3008240 & 0.7687531&  0.0027158 &\\
SuperWASP&    & 1.2987013 & 0.77     &  0.0042    &\\
TESS  &       & 1.2979596 & 0.77044  &  0.00009   &\\
  \noalign{\smallskip}\hline
\end{tabular}
\tablecomments{0.86\textwidth}{This table summarizes the results of the rotational periods of the HD\,HD345439 from the different datasets. The first column listed the names of the surveys; the filters that were adopted by the surveys are listed in the second column, and the rotational frequencies are listed in the third column. The rotation periods and corresponding errors are listed in columns 4 and 5. The sources of the data are listed in the last column.}
\end{table}

We extract the rotational frequencies from the NOWT photometric data in each band, which are denoted as F0 in Table\;\ref{tab:frequency}. The corresponding rotational periods of each band are listed in the fourth column of the table. The photometry data of each band exhibits a clear signature at around 0.7691 $\pm$ 0.0025 day; however, the signal-to-noise ratio (SNR) in the B band is low. The consistency of all observations indicates that the rotational period of HD 345439 is $0.7699\pm0.0014$ day, which is consistent with the value of $0.7701\pm0.0001$ derived by \cite{2015ApJ...811L..26W} and the value of $0.77018\pm0.00002$ day derived by \cite{2017MNRAS.467L..81H}.

\subsection{Pulsating behavior}
We utilized the PERIOD04 software\;\citep{2005CoAst.146...53L} to analyze the TESS and NOWT photometric data for investigating the pulsating behavior of HD\;345439. The Nyquist frequency of the TESS observations is $f\rm_{N}=360\rm\;d^{-1}$; thus, the frequencies in the range of $0<f<360\rm\;d^{-1}$ were removed from our analysis. We calculate the resolution frequency $f\rm_{res}=1/T\,$(T represents the duration of the observations) and use this information to distinguish between two contiguous frequencies. If the difference between two contiguous frequencies is less than the $f\rm_{res}$, we consider them to be unresolved. The rectified light curve is fitted using the following formula:
\begin{equation}\label{eq1}
m=m_{0}+\sum_{i=1}^{N} A_{i} \sin \left(2 \pi\left(f_{i} t+\phi_{i}\right)\right)
\end{equation}
where $m\rm_{0}$ is the zero-point, $A_{i}$ is the amplitude, $f_{i}$ is the frequency, and $\phi_{i}$ is the corresponding phase. The multifrequency least-square fit of the light curve for all detected significant
frequencies is then performed using Eq.\ref{eq1} to obtain the solution for all frequencies. The following search is conducted with the residual data that is obtained by subtracting the above solution from the rectification data. We adopt SNR $\geq4$, which is the criterion suggested by \cite{1993A&A...271..482B}, to detect the significant frequencies.

\begin{figure}[ht]
\centering
\includegraphics[scale=0.4]{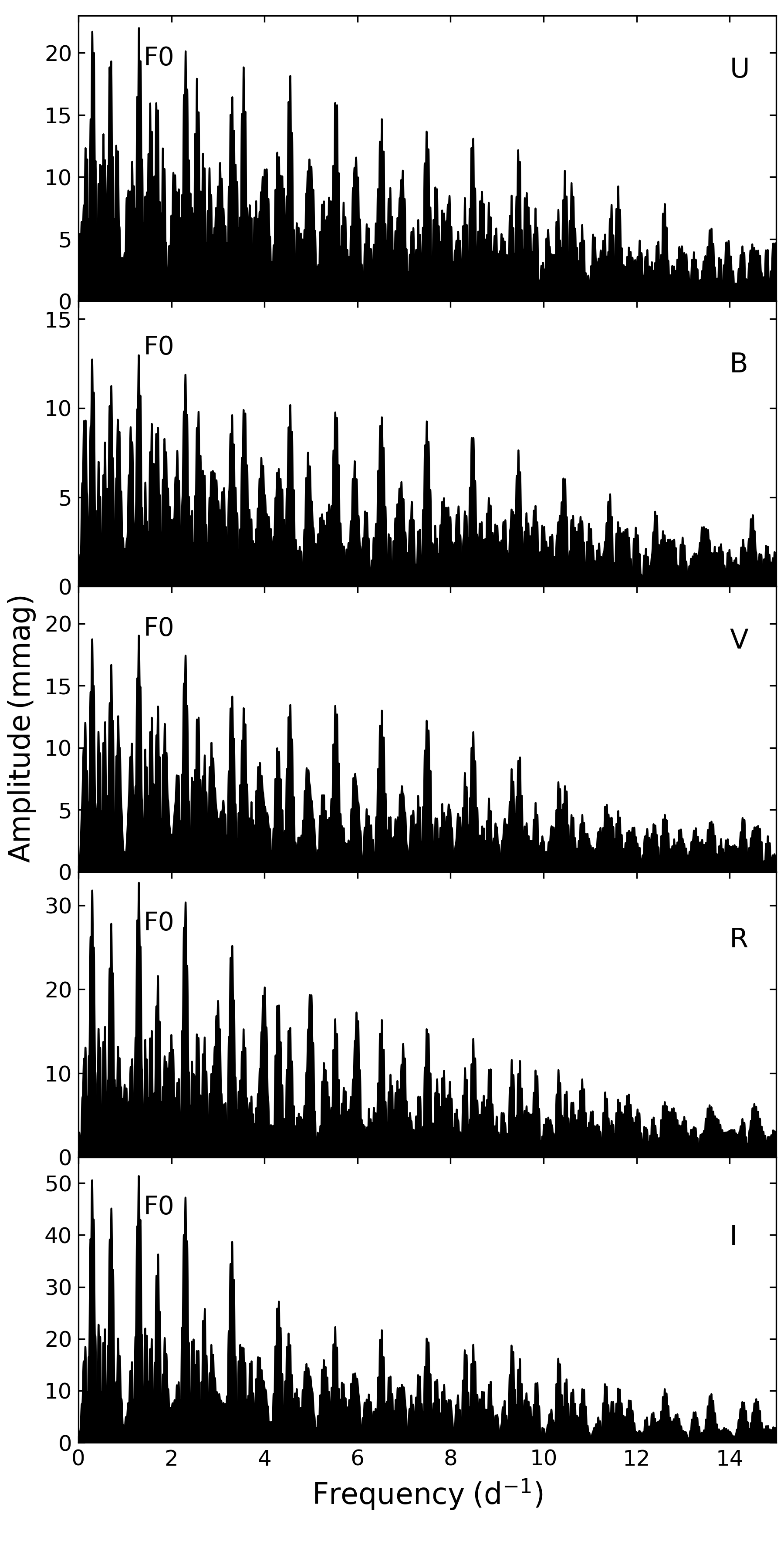}
\caption{Fourier amplitude spectra for multicolor NOWT light curve of HD\,345439. Top panel shows the Fourier amplitude spectrum of the $U$ band. The middle panels show the Fourier amplitude spectra of $B$, $V$, and $R$ band. The bottom panel shows the Fourier amplitude spectrum of the $I$ band. The rotational frequency\;is marked in each spectrum.}
\label{fig:nowt_frequency}
\end{figure}

We identify the significant frequencies in the multicolor NOWT photometry data; the Fourier amplitude spectrum of each band is shown in Fig.\;\ref{fig:nowt_frequency}. All significant frequencies are obtained according to the $f\rm_{res}=0.09\rm\;d^{-1}$ and SNR$\geq4$ restrictions, which are listed in Table\ref{tab:frequency}. We ﬁnd an independent frequency in the $U$ band. However, this independent frequency is not detected in the other bands, which indicates that it may be an instrumental artifact. Regarding the $B$, $V$, $R$, and $I$ bands, only the harmonics of the fundamental rotational frequency are detected.

\begin{table}%
\caption{Frequencies of multicolor NOWT Photometry}
\label{tab:frequency}
\begin{tabular}{ccccccc}
\hline\noalign{\smallskip}
\hline\noalign{\smallskip}
Filter& Frequency&  SNR& Period&  Period error& Identification&Amplitude\\
&  (day$^{-1}$) &  & (day) &(day) & & (mmag)\\
\hline\noalign{\smallskip}
U & 1.3017526 & 4.51914  & 0.7682053  & 0.0027970 &  F0 & 21.4193 \\
  & 3.5535679 & 4.81588  & 0.2814076  & 0.0002698 &  F1 & 16.4068 \\
\hline
B & 1.2991486 & 3.72138  & 0.7697400  & 0.0019904 &  F0 & 12.5016 \\
\hline
V & 1.3008506 & 4.69833  & 0.7687349  & 0.0023462 &  F0 & 18.6534 \\
\hline
R & 1.2988267 & 11.98561 & 0.7699334  & 0.0024346 &  F0 & 49.2253 \\
  & 3.9048166 & 4.20655  & 0.2560941  & 0.0002075 & 3F0 & 13.3104 \\
  & 6.5191430 & 4.05902  & 0.1533944  & 0.0000546 & 5F0 & 9.4123 \\
  & 2.6076572 & 8.56865  & 0.3834861  & 0.0001930 & 2F0 & 11.2453 \\
  & 6.4124358 & 5.42434  & 0.1559470  & 0.0000236 & 5F0 & 5.2630 \\
\hline
I & 1.3008240 & 6.91353  & 0.7687531  & 0.0027158 &  F0 & 31.7706 \\
  & 3.9675131 & 4.34097  & 0.2520472  & 0.0002174 & 3F0 & 14.8548 \\
  & 9.1107710 & 4.14031  & 0.1097602  & 0.0000100 & 7F0 & 3.4015 \\
  \noalign{\smallskip}\hline
\end{tabular}
\tablecomments{0.86\textwidth}{The results of searching the rotational period of HD345439 in multi-color photometry.}
\end{table}

\begin{figure}[ht]
\centering
\includegraphics[scale=0.4]{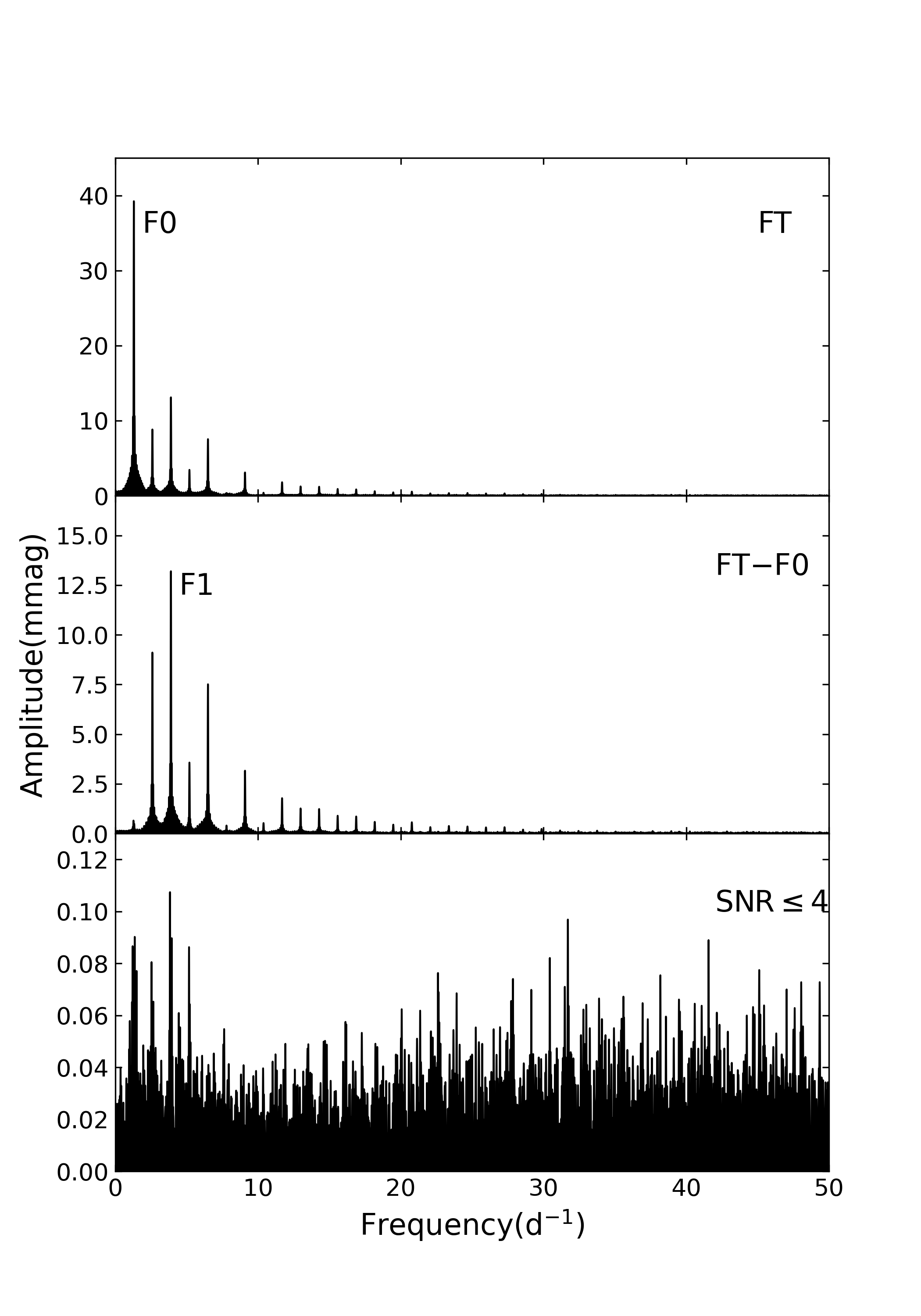}
\caption{Fourier amplitude spectra and the prewhiten process for the TESS light curve of HD\,345439. The top panel shows the rotational frequency F0, the middle panel shows the other harmonics of the rotational frequency, and the bottom panel shows the residual after subtracting all significant frequencies.}
\label{fig:frequency}
\end{figure}

The Fourier amplitude spectra of the TESS photometric data are shown in Fig.\;\ref{fig:frequency}. In addition to the rotational frequency, 39 significant frequencies are detected from the TESS photometric observations. Among these, 15 significant frequencies were removed after taking into account the $f\rm_{res}=0.04\rm\;d^{-1}$ restriction. The remaining frequencies are listed in Table \ref{tab:tess_frequency}; the corresponding SNR values, amplitudes, identifications, and periods are provided in columns 3--6, respectively. All the resolved significant frequencies are identified as the harmonics of the  rotational frequency, and no evidence of pulsating behavior is found in HD\;345439.

\begin{table}
\caption{All frequencies detected in photometric data of TESS\label{tab:tess_frequency}}
\begin{tabular}{cccccc}
\hline\noalign{\smallskip}
\hline\noalign{\smallskip}
No.&Frequency &  SNR &Amplitude &  Identification & Period\\
&(day$^{-1}$) &  & (mmag) & &(day)\\
\hline\noalign{\smallskip}
1& 1.29795232& 262.9869 & 39.2146409&  F0   & 0.770444336 \\
2& 3.89197588& 84.06976 & 13.2033909&  3F0  & 0.2569389   \\
3& 2.59590465& 170.52481& 9.08618912&  2F0  & 0.385222162 \\
4& 6.48788053& 160.28067& 7.54253258&  5F0  & 0.325243263 \\
5& 5.18992821& 113.87392& 3.35680069& 4F0   & 0.192680893 \\
6& 9.08378518& 151.64049& 3.22173686& 7F0   & 0.110086267 \\
7& 11.6796898& 69.02696 & 1.80845948& 9F0   & 0.085618712 \\
8& 12.9776422& 51.92077 & 1.28299003& 10F0  & 0.0770556   \\
9& 14.2755945& 54.0551  & 1.24464612& 11F0  & 0.070049622 \\
10& 15.5735468& 40.94505& 0.879632814& 12F0 & 0.064211449 \\
11& 16.8714991& 38.71091& 0.843725336& 13F0 & 0.059271556 \\
12& 18.1713325& 28.10908& 0.582071681& 14F0 & 0.055031737  \\
13& 20.7691183& 24.24462& 0.548933463& 16F0 & 0.048148409  \\
14& 10.3817375& 26.15445& 0.479181066& 8F0  & 0.09632299   \\
15& 19.4674038& 21.34688& 0.427989186& 15F0 & 0.051367918  \\
16& 3.91831115& 14.4029 & 0.462188854& 3F0  & 0.255211994  \\
17& 23.3631418& 14.63346& 0.377758953& 18F0 & 0.042802462  \\
18& 24.6648564& 13.82355& 0.336342688& 19F0 & 0.040543516  \\
19& 7.78771395& 16.50572& 0.332816778& 6F0  & 0.128407387  \\
20& 27.2626421& 11.7586 & 0.297142974& 21F0 & 0.036680231  \\
21& 22.0633084& 11.21728& 0.296931562& 17F0 & 0.045324118  \\
22& 25.9609276& 13.24085& 0.281336096& 20F0 & 0.038519425  \\
23& 29.8585467& 9.31929 & 0.231048733& 23F0 & 0.033491248  \\
24& 28.5568322& 7.19675 & 0.180087489& 22F0  & 0.03501789  \\
25& 33.7486415& 6.3148  & 0.158025325& 26F0  & 0.029630822  \\
  \noalign{\smallskip}\hline
\end{tabular}
\tablecomments{0.86\textwidth}{All frequencies that are detected in the photometric data of TESS are listed in the table. We calculated the corresponding period for each frequency, and identified their nature as independent frequencies, harmonics, or combinations.}
\end{table}

\begin{figure}
\centering
\includegraphics[scale=0.5]{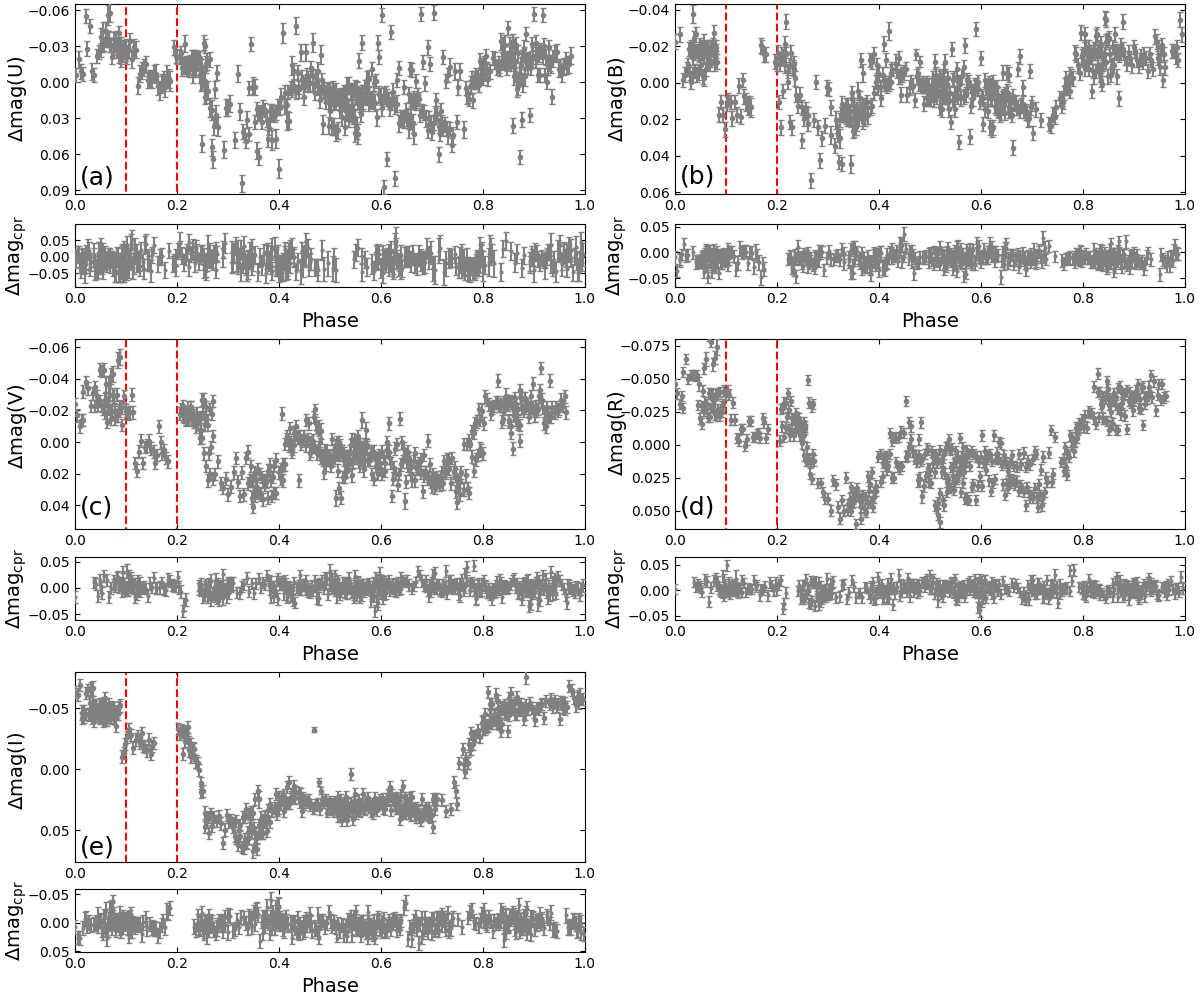}
\caption{Phase diagram of multicolor photometric observations with NOWT. A phase fold has been preformed for each band with the corresponding period. The light curves of the comparison star are shown under the light curves of HD\,345439. The vertical red dashed lines marked the parts of the light curves where the differences compared to the light curves were predicted by the RRM model.}
\label{fig:NOWT}
\end{figure}

\subsection{Morphology of the light curve}
The light curves of all new archival observations are shown in Fig. \ref{fig:periodogram}. We calculate the median magnitude in bins of 0.02-phase width and overlay these median values on the initial phase diagram to aid in the visual interpretation of the SuperWASP observations. The light curves show an analogical variation with a feature of the S-wave, and the maximum amplitude$\;\sim0.1\;$mag. The multicolor NOWT photometry is shown in Fig.~\ref{fig:NOWT}. The profiles and amplitudes of the light curves are approximately consistent with the archival observations.

The shapes of the light curves of all photometric observations show a typical double S-wave feature. The light curve obtained from the TESS observations exhibits a pronounced asymmetry between the primary and secondary eclipses. The same asymmetric components are detected in the multicolor observations. An unequal spacing between the primary and secondary eclipses is visible in each band, and the secondary eclipse is delayed with respect to the primary one by a phase of $\sim$ 0.4. The presence of these unequal spacings constitutes a strict constraint for the shape and density proﬁle of the magnetosphere clouds as well as a restriction on the magnetic obliquity angle {$\beta$} and the rotational inclination angle $i$. On the other hand, the multicolor NOWT photometry light curves show a different trend in the phase 0.1 to 0.2, compared with that of other photometric observations, and the amplitude variation decreases with the increase in the central wavelength of the filter. This feature region is marked between two vertical red dashed lines in Fig.~\ref{fig:NOWT}. The same trend does not appear in the other archival observations. Regarding the ASAS-SN and SuperWASP observations, it is the low time resolution that causes the absence of this amplitude variation. The bandpass of the TESS detector ranges from 600 to 1000 nm, and the central wavelength is 786.5 nm\,\citep{2015JATIS...1a4003R}. However, the variability of the multicolor NOWT photometry at the 0.15 phase is considerably reduced in the $I$ band; therefore, the amplitude variation is unobservable for the TESS detector. The amplitude variation of the multicolor light curves suggests that the material in the magnetic clouds may condense into larger particles, thereby causing an increase in the short-wavelength opacity.

\subsection{Centrifugal Breakout}
Being the archetype of helium-rich stars with large-scale magnetic fields, {$\sigma$} Ori E was conformed to have hard X-ray flares \cite[e.g.,][]{2004A&A...418..235G,2004A&A...421..715S}. \citet[hereafter referred to as TO05]{2005MNRAS.357..251T} speculated that the X-ray flares are associated with the centrifugal breakout and are universal in rapidly rotating magnetic early B star. Subsequently, \citet[hereafter referred to as UD06]{2006ApJ...640L.191U} demonstrated the feasibility of centrifugal breakout hypothesis via numerical magnetohydrodynamic\;(MHD)\;simulations.

\begin{figure}[ht]
\centering
\includegraphics[scale=0.38]{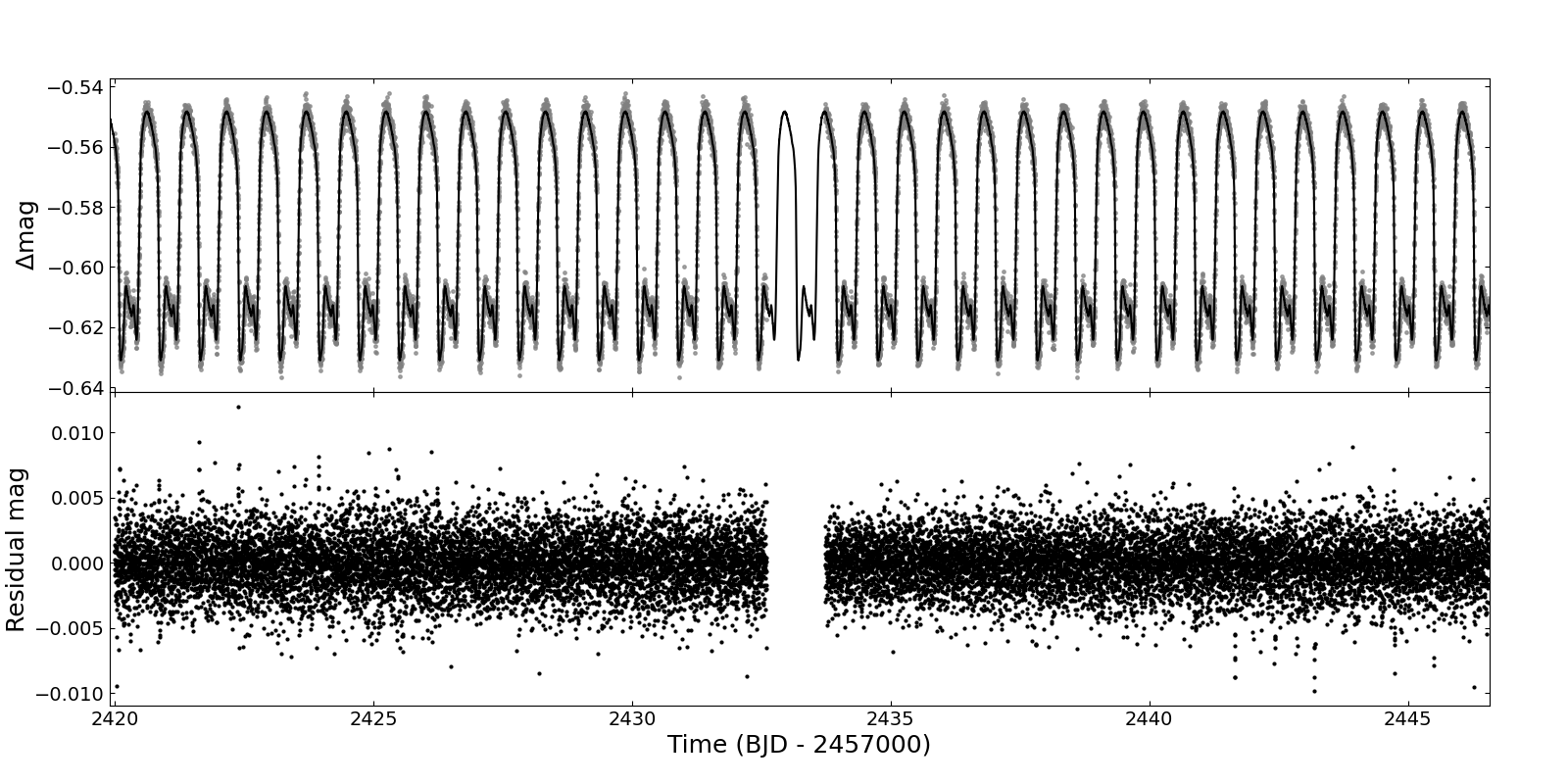}
\caption{Upper panel: prewhitened light curve of HD\;345439. Bottom panel: residual magnitudes after subtracting the light curve with the periodic signal.}
\label{fig:flare}
\end{figure}

In this study, we explore the existence of centrifugal breakout of HD\;345439 using the TESS photometric data. We use the functions in the LK package to prewhiten the TESS light curve for eliminating the instrumental variations, and the result is shown in the upper panel of Fig.~\ref{fig:flare}. After subtracting the light curve with the periodic signal, the residual magnitudes are shown in the bottom panel of Fig.~\ref{fig:flare}. As estimated by TO05 and UD06 works, the entire centrifugal breakout event would occur on a timescale of days. However, apart from the variation caused by the magnetosphere, the smoothed light curve shown in the bottom panel of Fig.~\ref{fig:flare} is relatively devoid of features. Indeed, the residual magnitudes do not vary significantly. Additionally, no systematic variations are observed at the extremum of the light curve. It can thus be inferred that no centrifugal breakout episodes occurred during the TESS photometric observations.

\subsection{Stellar Parameters}
The luminosity is derived by using the bolometric correction ($BC$), and the extinction coefficient$\;A_{V}$ is estimated from the intrinsic color$\;(B-V)\rm _{0}$. The distance modulus$\;(DM)\;$is evaluated using the distance from Gaia DR2. The Gaia DR2 distances are corrected using a weak distance prior that varies smoothly as a function of Galactic longitude and latitude according to a Galaxy model\,\citep{2018AJ....156...58B}. The intrinsic color$\;(B-V)\rm _{0}$ is determined from the smooth interpolation of the grids provided by \citet{2007ApJS..169...83L}. We use the relationship of$\;R(V)\equiv A(V)/E_{(B-V)}=3.1$ \citep{1989ApJ...345..245C} to derive the extinction coefficient$\;A_{V}$. The solar values, including $BC_{V,\odot}=-0.19$ mag, $M_{bol,\odot}=4.60$ mag, $M_{V,\odot}=4.83$ mag, and $V_{\odot}=-26.74$ mag were taken from a previous work \citep{2010AJ....140.1158T}. The absolute visual magnitude$\;(M_{V}=V-A_{V}-\rm DM$), the bolometric magnitude$\;(M_{bol}=M_{V}+BC)$, and the luminosity\;[$log(L/L_{\odot})=0.4\times(M_{bol,\odot}-M_{bol})$] are compiled in Table\;\ref{tab:parameters}.

\begin{figure}
\begin{minipage}[t]{0.25\linewidth}
\centering
\includegraphics[width=6in]{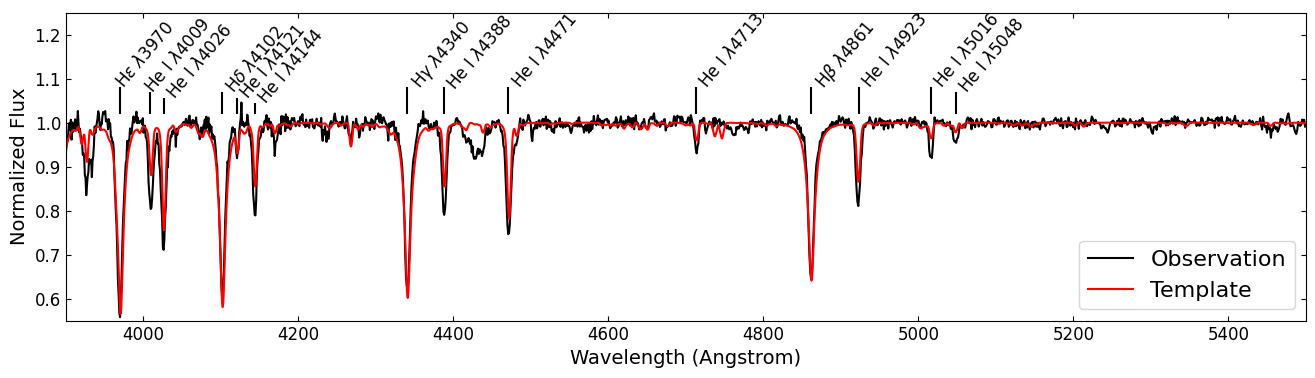}
\end{minipage}\\
\begin{minipage}[t]{0.35\linewidth}
\centering
\includegraphics[width=6.1in]{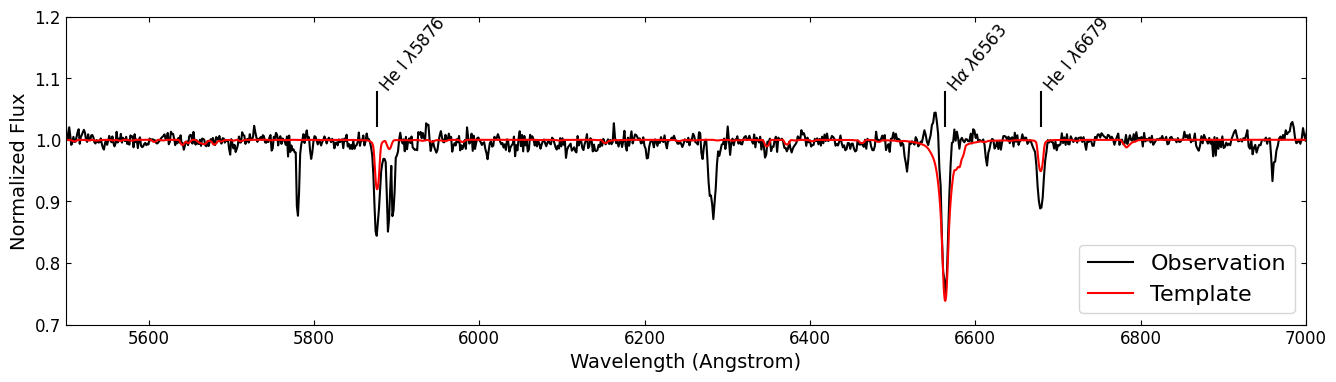}
\end{minipage}%
\caption{LAMOST-LRS spectrum of HD\;345439. The identified lines were marked in the panels. The black line represents the observed spectrum, and the red line represents the template spectrum. We truncate the wavelength at 700\,nm, and show the results in two panels.}
\label{fig:spectrum}
\end{figure}

We download the grids of ATLAS9 model atmospheres \citep{2003IAUS..210P.A20C} from the ATLAS websites\;\footnote{https://wwwuser.oats.inaf.it/castelli/grids.html}. The grids include the effective temperature ($T\rm_{eff}$) spanning a range of 15 000$\;\sim\;$35 000 K with a step of 1000\,K, the log$\,{g}$ in the range of 2.50$\;\sim\;$5.00 dex with a step of 0.50 dex, and the metal abundance in the range of -4.00$\;\sim\;$5.00 dex. The SPECTRUM program \citep{1994AJ....107..742G} was adopted to synthesize the spectral templates. The unbroadened infinite-resolution spectra were transformed into low resolution\;($R\sim1800$)\; spectra to be comparable with the LAMOST-LRS. To ensure the accuracy of analysis, we removed the spectral region with wavelength below 390\,nm, which is affected by instrumental ripples, and truncated the spectrum at 700\,nm. For determining the stellar atmospheric parameters, we matched the LAMOST-LRS spectrum with templates using a methodology similar to LSP3. The LSP3 adopts a cross-correlation algorithm to determine stellar radial velocity and uses the weighted means of parameters of the best-matching templates and values yielded by $\chi^{2}$ minimization \citep{2015MNRAS.448..822X}. The $\chi^{2}$ values calculated from the observing spectrum and matching template spectra are defined as:
\begin{equation}\label{eq2}
\chi^{2}=\sum_{i=1}^N \frac{(O_{i}-T_{i})^2}{\sigma_{i}^{2}}
\end{equation}
where $O_{i}$ is flux densities of the observing spectrum of the $i$th, $T_{i}$ is flux densities of the template spectra of the $i$th. N is the total pixel number used to calculate the $\chi^{2}$ values, and $\sigma_{i}$ is the error of flux density of the observing spectrum of the $i$th pixel.
As shown in Fig.\;\ref{fig:spectrum}, the best fitting result reproduces with the effective temperature at $22\pm1$\,kK, log$\,g=4.00\pm0.22$ and $[Fe/H]=-0.50$\;dex.

\begin{figure}[ht]
\centering
\includegraphics[scale=0.3]{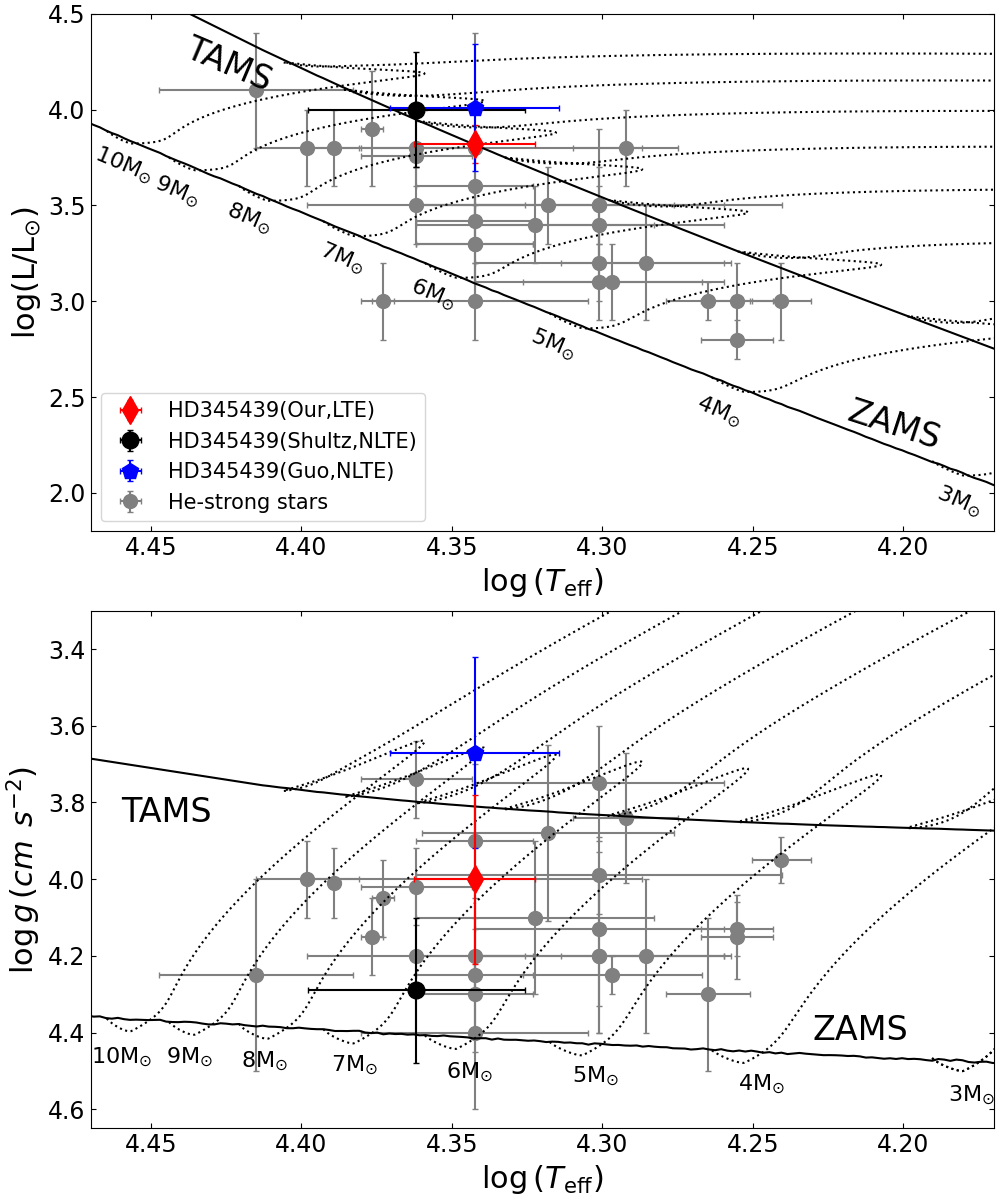}
\caption{Hertzsprung–Russell diagram (top panel) and Kiel diagram (bottom panel). The position of HD 345439 is marked with the results of both the LTE and NLTE. The red diamond indicates the result derived from the LTE model. The black dot and blue pentagon represent the results derived from the NLTE models. The locations of the He-rich stars are marked with gray dots. For HD 345439, the red diamond represents the result of this work, the blue pentagon represents the result of \cite{2021ApJS..257...54G}, and the black dot represents the result of \cite{2019MNRAS.485.1508S}. The parameters of other He-rich stars were obtained from \cite{2019MNRAS.485.1508S}}
\label{fig:keil_diagram}
\end{figure}

\begin{table}[ht]
\caption{Stellar parameters of HD 345439}
\label{tab:parameters}
\begin{tabular}{cccc}
\hline\noalign{\smallskip}
\hline\noalign{\smallskip}
Sp.\,type$^{1}$& B2V& & \\
$\pi$(mas)$^{2}$& $0.3739\pm0.0192$&&\\
$Dis.$(pc)$^{3}$& $2141.85_{-179.54}^{+214.64}$& & \\
\hline
Photometric& Our Work&Shultz$^{4}$ \\
\hline
$(B-V)_{0}$\,(mag)& -0.245                 &...           &\\
$E_{(B-V)}$\,(mag)& $0.745\pm0.016$        &...           &\\
$A_{V}$\,(mag)    & $2.31\pm 0.05$         &$2.2\pm0.1$   &\\
$BC$\,(mag)       & $-2.11\pm 0.11$        &$-2.4\pm0.4$  &\\
$M_{V}$\,(mag)    & $-2.85\pm 0.16$        &$-2.9\pm0.3$  &\\
$M_{bol}$\,(mag)  & $-4.96\pm 0.25$        &$-5.3\pm0.7$  &\\
$P\rm_{rot}$\,(day) & $0.7699\pm 0.0014$     &              &\\
\hline
Physical                &                         &                         & Guo$^{5}$             \\
\hline
$T\rm_{eff}$\,(kK)        &  $22\pm1^{L}$               & $23\pm2^{N}$                &$22\pm1.5^{N}$             \\
log$g$\,(cgs)             &  $4.00\pm0.22^{L}$          & $4.29\pm0.19^{N}$           & $3.67\pm0.25^{N}$          \\
log($L/L_{\odot}$)   &  $3.82\pm0.1$           & $4.0\pm0.3$             & $4.01_{-0.17}^{+0.33}$ \\
$[Fe\slash H]$(dex)          &  -0.5                   & ...                     & -0.35                  \\
$M/M_{\odot}$        &  $8.00_{-0.51}^{+0.50}$ & $8.99_{-1.19}^{+1.02}$  & $8.59_{-1.00}^{+2.00}$ \\
$R/R_{\odot}$        &  $5.59\pm0.06$         & $6.29_{-1.83}^{+0.93}$  & $7.00_{-1.30}^{+3.21}$ \\
$\tau_{age}$\,(Myr)       &  $26.79_{-6.79}^{+7.03}$& $22.50_{-5.23}^{+0.09}$ & $26.52_{-7.78}^{+6.60}$\\
  \noalign{\smallskip}\hline
\end{tabular}
\tablecomments{0.86\textwidth}{References: $^{1}$\cite{2014ApJ...784L..30E}, $^{2}$\cite{2020yCat.1350....0G}, $^{3}$\cite{2018AJ....156...58B}, $^{4}$\cite{2019MNRAS.485.1508S}, $^{5}$\cite{2021ApJS..257...54G}. $^{N}$ represents the parameter determined from NLTE models, $^{L}$ represents the parameter determined from LTE models.}
\end{table}

A grid of evolution tracks with a mass ranging from 4\,M$_{\odot}$ to 10\,M$_{\odot}$ with a step of 0.02\,M$_{\odot}$ was generated using the Modules for Experiments in Stellar Astrophysics (MESA)\,\citep{Paxton2011,Paxton2013, Paxton2015,Paxton2018,Paxton2019}. The rotational effect of the star having $v/v\rm_{crit}$ = 0.75, a restriction which is imposed by photometric observations, was used in the grid. The luminosity, mass, and age of HD\;345439 were estimated from its location on the Hertzsprung$-$Russell Diagram (HRD) with the evolutionary tracks. The positions of HD\;345439 on the Kiel-Diagram\,(KD) and HRD are shown in Fig.\ref{fig:keil_diagram}. The mass, luminosity, and age were derived as $\;M= 8.00\,_{-0.51}^{+0.50}\rm\,M_{\odot}$, log$(L/L_{\odot})=3.82\pm0.1$\,dex, and $\tau\rm_{age}= 26.79\,_{-6.79}^{+7.03}$\,Myr, respectively. According to the Stefan–Boltzmann law, the radius was estimated to be $5.59\pm0.06\rm\,R_{\odot}$.

\section{DISCUSSION}
\subsection{Parameters restriction under the RRM model}
Following a previous report\;\citep{1985Ap&SS.116..285N}, the TO05 work developed an RRM model to interpret the interaction of the magnetosphere with other phenomena in hot stars and indicated that magnetospheric clouds are the primary factor responsible for the variation in the light curve of rapidly rotating magnetic stars.

%The parameters, including the magnetic obliquity angle {$\beta$} and the rotational inclination angle $i$ were taken from spectropolarimetry and high-resolution spectroscopy \citep{2017MNRAS.467L..81H, 2020MNRAS.499.5379S}.
\begin{figure}[ht]
\centering
\includegraphics[scale=0.2]{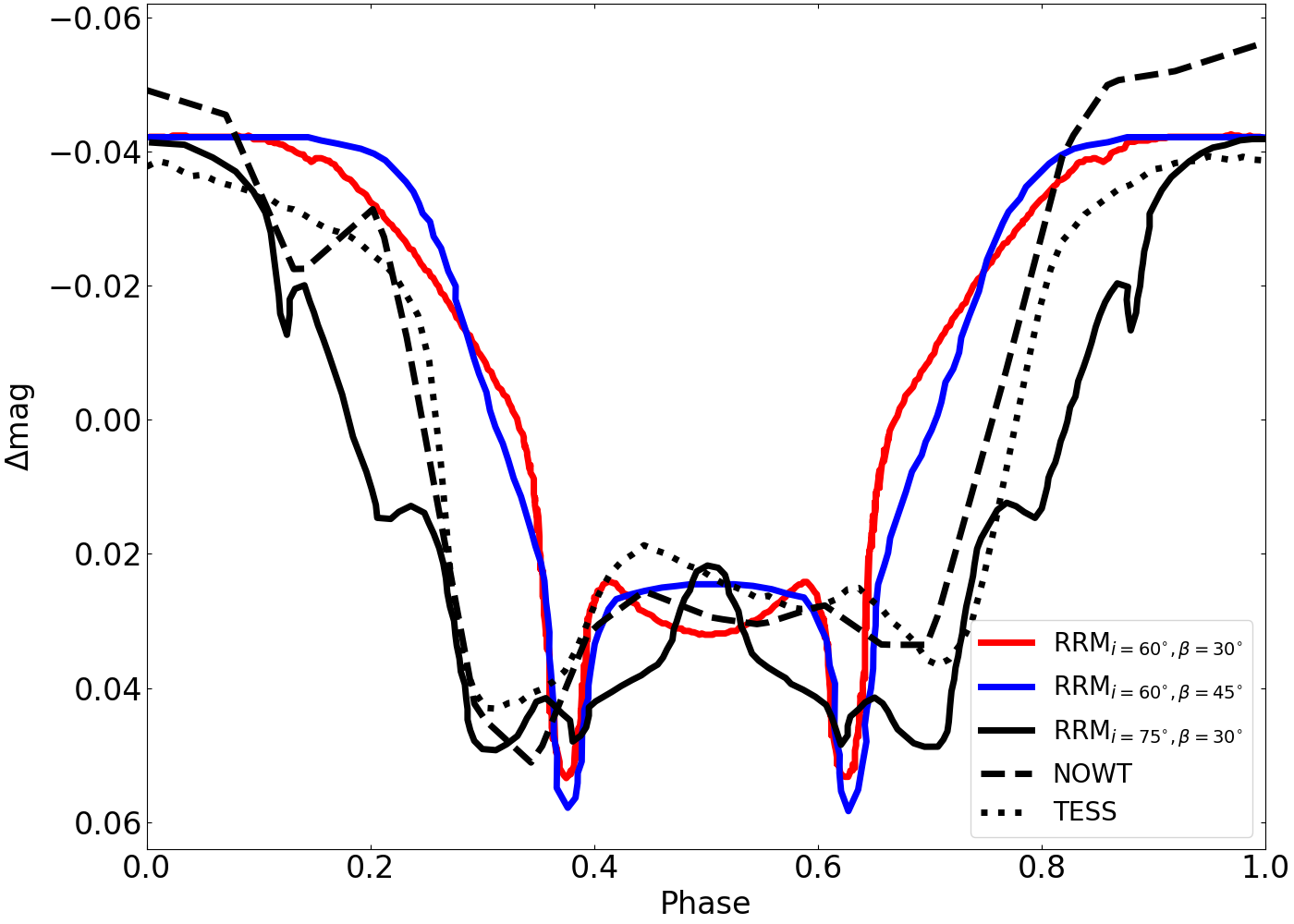}
\caption{Light curves derived from the photometric observations and synthesized with the RRM model. The bule line represents the light curve that was synthesized with the RRM model using the parameters taken from \cite{2020MNRAS.499.5379S}. The black line represents the light curve that was synthesized with the RRM model using the parameters taken from \cite{2017MNRAS.467L..81H}. We adjusted the parameters to reproduce the variations at the 0.5 phase (red line). The Dotted and dashed lines represent the TESS and NOWT photometric observations, respectively.}
\label{fig:lc_shape}
\end{figure}
By applying the light-curve synthesis demonstrated by the UD06 work\footnote{http://www.astro.wisc.edu/~townsend/static.php?ref=rrm-movies}, we obtained the shapes of light curves based on the RRM model. Since the light-curve synthesis does not provide the absolute flux, we used the normalized flux converted to the corresponding $\Delta$mag. To generate the theoretical light curves, we used a grid with the rotational factor\;($v/v\rm_{crit}$)\;in the range from 0.00 to 0.99 with a step of 0.25, the magnetic obliquity angle {$\beta$} in the range of 0\,$^{\circ}$$\sim$90\,$^{\circ}$ with a step of 15\,$^{\circ}$, and the rotational inclination angle $i$ in the range of 0\,$^{\circ}$$\sim$90\,$^{\circ}$ with a step of 15\,$^{\circ}$. For the accuracy of the analysis, we used the light curves obtained from the TESS observations and the multicolor NOWT photometry. The $\chi^{2}$-minimum method was used to estimate the difference between the RRM model and the observations. Fig.\;\ref{fig:lc_shape} shows the comparison between the photometry light curves and the light curves synthesized using the RRM model. The best fit\,(black line) indicates that we can restrict the magnetic obliquity angle {$\beta$} and the rotational inclination angle $i$ to the values satisfying the approximate relation {$\beta$ + $i$ $\approx$ 105\,$^{\circ}$}\,($i=$75\,$^{\circ}$ and $\beta$=30\,$^{\circ}$)
Based on the work by \cite{2020MNRAS.499.5379S}, we also adopt the assumptions that $i$=60\,$^{\circ}$ and $\beta$=45\,$^{\circ}$. However, there are several differences between the simulated result and photometric observations. The unequal eclipses may be caused by the mismatch parameters with the magnetic obliquity angle {$\beta$} and the rotational inclination angle $i$ \citep{2005ApJ...630L..81T}. Our result is more consistent with that report by \cite{2017MNRAS.467L..81H}.

Although we are able to reproduce the light curve for HD\;345439, there are evident discrepancies between the theoretical and observational light curves. In contrast to the synthesized light curves, asymmetrical depths of eclipses are detected in the observational light curves. \cite{2005ApJ...630L..81T} pointed out that this asymmetry in the eclipses may be caused by an offset between the center of the dipole magnetic fields and the rotation axis. Furthermore, in the phase range of $0.3\sim0.7$, the observational and synthesized light curves show opposite trends. One hypothesis that could explain this discrepancy is that the magnetosphere clouds of HD\;345439 have more complex geometries, as was also reported by \cite{2013ApJ...766L...9C}. In the $0.3\sim0.7$ phase, the magnetospheric column density is affected by other small scale magnetic fields, which results in larger density values than those estimated by the RRM model. The TO05 work presented the morphology of magnetic clouds with pure poloidal magnetic fields. Nevertheless, the MHD simulations show that in order to maintain stable global scale magnetic fields on long timescale, the ratio of the energies of toroidal and poloidal components of the magnetic fields must have a specific value\,\citep{2009MNRAS.397..763B, 2010ApJ...724L..34D}. On the other hand, the opposite trend may be caused by chemical anomalies and starspots, which are from microscopic chemical transport processes. It is thus necessary to determine the surface chemical spot structure using Doppler imaging (DI) to further restrict the light curve of HD\,345439\,\cite[e.g.,][]{2022MNRAS.510.5821K}.
\subsection{Mass leakage process}
No evidence of centrifugal breakout was found in the TESS photometry. Several factors could explain the failure of the RRM model in detecting the centrifugal breakout. One of the reasons is that no relevant scheduling of the TESS run was suitable for searching for the ﬂare of the centrifugal breakout. Another factor could be that the mass of the magnetic clouds does not reach the crucial value necessary to generate the centrifugal breakout. The TO05 work predicted that the crucial mass was dependent strongly on the magnetic strength, stellar radius, mass-loss rate, and escape velocity. HD\;345439 has a more extreme state than {$\sigma$} Ori E; thus, its breakout timescale is larger than the latter with a low incidence of breakout episode.

Although the centrifugal breakout is a natural result in MHD for rapidly rotating magnetic stars \citep{2006ApJ...640L.191U}, no centrifugal breakout event was detected for HD\;345439. A uniform situation occurred in $\sigma$ Ori E, which indicates the low incidence of events \citep{2013ApJ...769...33T}. In addition to the centrifugal breakout, the ongoing reductions in the magnetospheric column density might indicate the mass leakage of the magnetosphere. \cite{ 2018MNRAS.474.3090O} presented diffusion-plus-drift magnetospheres (DDM) model to interpret the mass leakage of magnetic hot stars. In this scenario, plasma diffuses in both directions and escapes from the star via drift. The mass leakage process in the DDM model is more gradual than in the RRM model, and in particular the centrifugal breakout does not occur in the DDM model. \cite{2020MNRAS.499.5379S} analyzed the H$\alpha$ line profiles of the magnetic B-type stars, and pointed out that the H$\alpha$ emission and strength are independent of
the effective temperature, luminosity, and mass-loss rate; these results support the hypothesis of the occurrence of centrifugal breakout events. However, \cite{2020MNRAS.499.5379S} also reported that centrifugal breakouts must be a continuous process, and the H$\alpha$ emission is in contradiction with the model predictions. The detection of the centrifugal breakout using the TESS data presented in this work supports the results of \cite{2020MNRAS.499.5379S}.

\subsection{Discrepancies in the stellar parameters}
We obtained the stellar parameters of HD\;345439 from the LAMOST-LRS spectrum. The basic parameters were also measured by previous works: \cite{2021ApJS..257...54G} derived $T\rm_{eff}=22\pm1.5$\,kK, log$\,g=3.67 \pm0.25$, and $[Fe/H]=-0.35$\,dex with the Stellar LAbel Machine\,(SLAM). The SLAM \citep{2020ApJS..246....9Z} is a data-driven method based on support vector regression (SVR) which is a robust nonlinear regression technique and often used to build the mapping from stellar spectra to stellar labels in previous works \citep[e.g.,][]{2015MNRAS.447..256B,2015MNRAS.452.1394L}. \cite{2020MNRAS.499.5379S} determined $T\rm_{eff}=23\pm2$\,kK, log$\,g=4.29\pm0.19$, and log$(L/L_{\odot})=4.0\pm0.3$\,dex from the Gaia photometric observations and high-resolution NLTE atmosphere spectra. Regarding the LAMOST-LRS spectrum, it should be noted that the lowest value of the surface gravity was obtained from the NLTE atmosphere using the SLAM, while the highest value was from the LTE atmosphere. A similar discrepancy occurred for other hot stars \citep[e.g.,][]{2021MNRAS.507.1283S}. As one of the main reasons, we adopted a different atmospheric model, which indicated the different line list, model atoms, and opacity will be used to generate final template spectra. Secondly, there are differences in the method of how to derive the stellar parameters of the HD\;345439. Different grids were used between our study and \cite{2021ApJS..257...54G} to increase the difference in the surface gravity. The surface gravity and effective temperature derived in the present work are consistent with the results of \cite{2020MNRAS.499.5379S}.

The locations of the HD\;345439 derived from the NLTE and LTE atmospheres are marked in Fig.\;\ref{fig:keil_diagram}. The mass, radius, and age were also derived using the evolution tracks in the HRD, and the final results are listed in Table\;\ref{tab:parameters}. Being a magnetic He-rich star, HD\;345439 is located in the colony of other main-sequence\;(MS)\;magnetic He-rich stars on the HRD and KD. HD\;345439 is around the terminal-age main sequence (TAMS), which suggests that HD\;345439 is an old B-type star. HD\;345439 has a larger surface gravity than that inferred from its luminosity derived from the NLTE or LTE atmosphere. The other magnetic He-strong early-type stars are also marked in the HRD and KD. The locations of the He-strong stars in the HRD and KD are also characterized by systematic discrepancies. We also mark the other magnetic He-strong early-type stars on the HRD and KD. The locations of He-strong stars between the HRD and KD also have systematic discrepancies.

The surface gravity of magnetic early-type stars is usually inferred from the pressure broadening of some specific H and He lines \citep[e.g.,][]{2007A&A...468..263C,2015MNRAS.451.1928S}. Strong magnetic fields contribute to the magnetic pressure via the Lorentz force, which results in the ﬂux of the spectral lines changing by several percent. Additionally, the magnetic clouds reduce this flux by approximately 2\%\,\citep{2004A&A...420..993V,2014Ap&SS.352...95V, 2019MNRAS.490..274S,2020MNRAS.499.5379S}. The flux variations result in a larger value of the surface gravity inferred from the template, and it can be seen that 0.2 dex variations in the surface gravity result in a mass uncertainty as high as $\sim20\%$. On the other hand, $\tau\rm_{age}$ of HD\;345439 has $\sim30\%$ uncertainty caused by the uncertainty in both log$(L/L_{\odot})$ and $T\rm_{eff}$. Although the number of magnetic stars are growing, only a dozen rapidly rotating magnetic stars have been conﬁrmed at the moment. It is important to obtain a detailed characterization of all known samples for understanding the evolution of rapidly rotating magnetic stars. Thus, more future works are required to perform further photometric observations of rapidly rotating magnetic stars and search for more samples for this type of star \citep[e.g.,][]{2006A&A...450..777B,2013A&A...551A..33H}.

\section{SUMMARY}
Based on the photometric data obtained from the NOWT observations and new archives, we calculated the rotational period of HD\;345439 to be $0.7699\,\pm\,0.0014$\,days. Our results are consistent with those of \cite{2015ApJ...811L..26W} and \cite{2017MNRAS.467L..81H}. In addition, no pulsating behavior was detected in HD\;345439. We compared the light curves obtained from the photometric observations with those synthesized under the assumptions of the RRM model. The asymmetric components in the light curves of HD\;345439 may be caused by an offset between the center of the dipolar magnetic ﬁelds and the rotation axis. The unequal spacing of the eclipses was restricted by the relation $\beta + i \thickapprox 105^{\circ}$. No centrifugal breakout events were detected in the photometric observations possibly because of the lack of an adequate scheduling of the TESS or the fact that the magnetic clouds could not reach the critical mass. We derived the stellar parameters from the LAMOST-LRS spectrum and compared them with the results obtained from the NLTE atmosphere.  The obtained stellar parameters are consistent with those reported by \cite{2019MNRAS.485.1508S} while they are in slight disagreement with those of \cite{2021ApJS..257...54G}. These discrepancies are caused by the differences between the NLTE and LTE atmospheres as well as the method adopted to derive the physical parameters. On the other hand, the SLAM can only derive the local optimal solution for the LAMOST-LRS spectrum of HD\;345439.
\normalem
\begin{acknowledgements}
This work received the generous support of the National Natural Science Foundation of China under grants U2031204, 12163005, 11863005, and the Natural Science Foundation of Xinjiang No. 2021D01C075. We gratefully acknowledge the science research grants from the China Manned Space Project with Nos. CMS-CSST-2021-A08 and CMS-CSST-2021-A10. J.Z.L. acknowledges the science
research grants from the Ministry of S\&T No. SQ2022YFE010103 Guoshoujing Telescope (the Large Sky Area Multi-Object Fiber Spectroscopic Telescope LAMOST) is a National Major Scientific Project built by the Chinese Academy of Sciences. Funding for the project has been provided by the National Development and Reform Commission. LAMOST is operated and managed by the National Astronomical Observatories, Chinese Academy of Sciences. This paper includes data collected with the TESS mission, obtained from the MAST data archive at the Space Telescope Science Institute (STScI). Funding for the TESS mission is provided by the NASA Explorer Program. STScI is operated by the Association of Universities for Research in Astronomy, Inc., under NASA contract NAS 5–26555.
\end{acknowledgements}

\bibliographystyle{raa}
\bibliography{bibtex}
\end{document}